\documentclass[11pt,a4paper]{article}
\usepackage[utf8]{inputenc}
\usepackage[T1]{fontenc}
\usepackage{lmodern}
\usepackage{microtype}
\usepackage[a4paper,margin=25mm]{geometry}
\usepackage{xcolor}
\usepackage{pdfpages}
\usepackage{fancyhdr}
\usepackage{hyperref}

\definecolor{locogreen}{RGB}{35,92,72}
\definecolor{locogray}{RGB}{80,80,80}

\hypersetup{
  colorlinks=true,
  linkcolor=locogreen,
  urlcolor=locogreen,
  pdftitle={LOCO 2026 Lightning Talk Abstracts: 2nd International Workshop on Low Carbon Computing},
  pdfauthor={Ignatius Ezeani; Adrian Friday; Abdessalam Elhabbash; John Vidler; Daniel King; Paul Dempster},
  pdfsubject={Collected lightning-talk contributions from LOCO 2026},
  pdfkeywords={low carbon computing, sustainable computing, green computing, carbon-aware computing, sustainable AI}
}

\newcommand{\contrib}[5]{%
  \noindent\begin{minipage}{\textwidth}
  \textbf{#1}\hfill\textcolor{locogray}{p.~\pageref{#5}}\\[-0.1em]
  {\small\textbf{#2}}\\[-0.1em]
  {\small #3}\\[-0.1em]
  {\footnotesize\textcolor{locogray}{#4}}
  \end{minipage}\par\vspace{0.9em}
}

\begin{document}

\begin{titlepage}
\thispagestyle{empty}
\vspace*{1.8cm}
{\color{locogreen}\rule{\textwidth}{1.5pt}}\\[1.2cm]
{\Large\bfseries 2nd International Workshop on Low Carbon Computing}\\[0.45cm]
{\fontsize{30}{35}\selectfont\bfseries LOCO 2026\\[0.15cm] Lightning Talk Abstracts}\\[0.7cm]
{\Large Collected proceedings volume}\\[1.3cm]

{\large 10--11 September 2026}\\[0.15cm]
{\large Lancaster University, Lancaster, United Kingdom}\\[1.4cm]

{\bfseries Edited by}\\[0.3cm]
Ignatius Ezeani, Adrian Friday, Abdessalam Elhabbash,\\
John Vidler, Daniel King, and Paul Dempster\\[1.4cm]

\begin{minipage}{0.88\textwidth}
\small
This volume collects fourteen accepted lightning-talk contributions from LOCO 2026. The individual contributions are reproduced in their camera-ready form and retain their original authorship, affiliations, references, and internal formatting.
\end{minipage}

\vfill
{\small Main LOCO 2026 proceedings index: \href{https://arxiv.org/abs/2608.02072}{arXiv:2608.02072}}\\[0.5cm]
{\color{locogreen}\rule{\textwidth}{1.5pt}}
\end{titlepage}

\section*{Editorial introduction}
\addcontentsline{toc}{section}{Editorial introduction}
\begingroup\small

The 2nd International Workshop on Low Carbon Computing (LOCO 2026) brings together researchers and practitioners examining how computing can reduce its greenhouse gas emissions and wider environmental impacts. The workshop spans technical, methodological, organisational, and critical perspectives, reflecting the fact that lower-carbon computing cannot be achieved through hardware or software efficiency alone.

The lightning-talk track was designed for concise contributions that can stimulate discussion around emerging results, work in progress, practical interventions, methodological proposals, and critical perspectives. This collected volume brings together fourteen such contributions accepted for presentation at LOCO 2026. The contributions were reviewed through the LOCO 2026 programme process and are presented here as a single archival collection.

Across the volume, five connected themes emerge. First, several contributions examine how environmental impacts can be made visible and actionable. The E-SCOUT study asks whether carbon reporting changes researcher behaviour (L01), while the Green Algorithms Dashboard develops user-facing, comparable energy and carbon reporting across digital research infrastructures (L02). Carbon.txt and DIST address a related transparency problem at the level of web services and sustainability claims (L04).

Second, the collection considers when and how computation should be scheduled to reduce operational emissions. Carbon-aware workflow scheduling under deadlines explores the trade-offs created by precedence constraints, heterogeneous platforms, and time-varying energy supply (L03). The CAML-TC contribution extends this concern to machine-learning workloads, combining carbon-intensity forecasting with heuristic and reinforcement-learning scheduling (L14). Green TTA approaches the same broad problem from inference-time adaptation, asking how accuracy gains should be evaluated alongside deployment energy use (L06).

Third, several papers focus on the sustainability of computational methods and research infrastructure. Green BOA studies the environmental break-even point of machine-learning-based data compression (L05), while sustainability-aware GEMM benchmarking explores performance, energy, and carbon trade-offs across CPUs, GPUs, and FPGAs (L09). These papers highlight a recurring LOCO question: when does additional computation genuinely reduce system-level resource use, and when does it simply shift impacts elsewhere?

Fourth, the volume moves beyond efficiency towards sufficiency, longevity, and alternative operational models. Computational sufficiency in land-surface modelling asks how research communities determine how much computation is enough (L07). Interoperability is proposed as a carbon-reduction strategy by extending the useful life of hardware (L08), while software-centric measures against premature hardware obsolescence examine how software ecosystems themselves can drive unnecessary replacement (L13). A solar-powered Mastodon instance demonstrates a deliberately constrained model in which planned downtime becomes part of the service design (L10).

Finally, two contributions question the assumptions that underpin continued growth in computational demand. Counting the Stones in My Computer examines the material and temporal realities embedded in microchips and computational speed (L11). Towards Post-Growth AI studies the motivations behind AI use and asks how environmental awareness may interact with patterns of consumption and expectations of technological performance (L12).

Taken together, these fourteen contributions illustrate the breadth of low-carbon computing as a research agenda. They also show why the field benefits from bringing engineering approaches into conversation with behavioural research, institutional change, sufficiency, materiality, and post-growth perspectives. The collection is intended to preserve these concise contributions while making their relationships visible as part of the wider LOCO 2026 proceedings.

\textbf{Editorial note on formatting.} The contributed papers are reproduced from the authors' final camera-ready PDFs. Differences in typography, page size, spelling conventions, citation style, and internal layout therefore reflect the original submissions rather than editorial re-typesetting.
\endgroup

\clearpage
\section*{Contents}
\addcontentsline{toc}{section}{Contents}

\contrib{LOCO2026/L01}{Does measuring carbon change behaviour? Designing a multi-centre trial in research organisations}{Christina Bremer, Jan van der Scheer, Lo\"ic Lannelongue}{Behaviour change; carbon reporting; scientific computing}{l01}
\contrib{LOCO2026/L02}{A unified user-facing carbon and energy monitoring dashboard for digital research infrastructures}{Navirah Kamal, Laurent Gil, Andrew Turner, Lo\"ic Lannelongue}{Carbon and energy monitoring; digital research infrastructure}{l02}
\contrib{LOCO2026/L03}{Carbon-Aware Workflow Scheduling Under Deadlines}{Dominik Schweisgut}{Carbon-aware scheduling; workflows; deadlines}{l03}
\contrib{LOCO2026/L04}{Making transparent, verifiable sustainability claims on the web with carbon.txt and DIST}{Tim Cowlishaw, Hannah Smith, Chris Adams, Fershad Irani}{Sustainability claims; web standards; transparency}{l04}
\contrib{LOCO2026/L05}{Green BOA: Determining the environmental break-even point for ML-based data compression}{Caterina Doglioni, Akshat Gupta, Thomas Elliott, Hanzila Hussain, Sanjiban Sengupta, Zhengkai Sun}{Data compression; machine learning; environmental break-even}{l05}
\contrib{LOCO2026/L06}{Green TTA: Benchmarking the Energy Efficiency of Test-Time Adaptation Methods}{Vivian White, Evan Shelhamer}{Test-time adaptation; energy efficiency; Green AI}{l06}
\contrib{LOCO2026/L07}{Ambition or excess? Computational sufficiency in land surface modelling}{Joe Marsh Rossney, Carolynne Lord, Gordon S. Blair, Lily Gouldsbrough, Marcia Tavares Smith, Kelly Widdicks}{Computational sufficiency; land-surface modelling; research infrastructure}{l07}
\contrib{LOCO2026/L08}{Interoperability as a Carbon Reduction Strategy}{Kern Tallett}{Interoperability; hardware longevity; embodied carbon}{l08}
\contrib{LOCO2026/L09}{Towards Sustainability-Aware Benchmarking and Optimisation of Generalised Matrix-Matrix Multiplication (GEMM) Operations for CPUs, GPUs, and FPGAs on HPC Systems}{Ishaan Alidina, Sylvain Laizet}{GEMM; HPC; sustainability-aware benchmarking}{l09}
\contrib{LOCO2026/L10}{Small-scale solar-powered social media: power saving through planned downtime}{James Coxon}{Solar-powered computing; Mastodon; planned downtime}{l10}
\contrib{LOCO2026/L11}{COUNTING THE STONES IN MY COMPUTER: Unfolding the Chip Reality}{Coralie Gourguechon}{Materiality; microchips; computing with limits}{l11}
\contrib{LOCO2026/L12}{Towards Post-Growth AI: What Motivates its Usage?}{Lena Pohlmann, Hajo Boomgaarden, Sophie Lecheler, Emanuel Sallinger}{Post-growth AI; environmental awareness; AI use}{l12}
\contrib{LOCO2026/L13}{Software-centric measures against premature hardware obsolescence}{Stephen Kell}{Software sustainability; hardware longevity; obsolescence}{l13}
\contrib{LOCO2026/L14}{When to Train: A Carbon-Aware Scheduler for ML Workloads}{Sufiyan Ul Rehman}{Carbon-aware machine learning; scheduling; reinforcement learning}{l14}

\clearpage
\fancyfoot[L]{\footnotesize LOCO2026/L01 \quad LOCO 2026 Lightning Talk Abstracts}
\fancyfoot[R]{\footnotesize \thepage}
\includepdf[pages=1,scale=0.89,pagecommand={\thispagestyle{fancy}},addtotoc={1,section,1,{L01: Does measuring carbon change behaviour?},l01}]{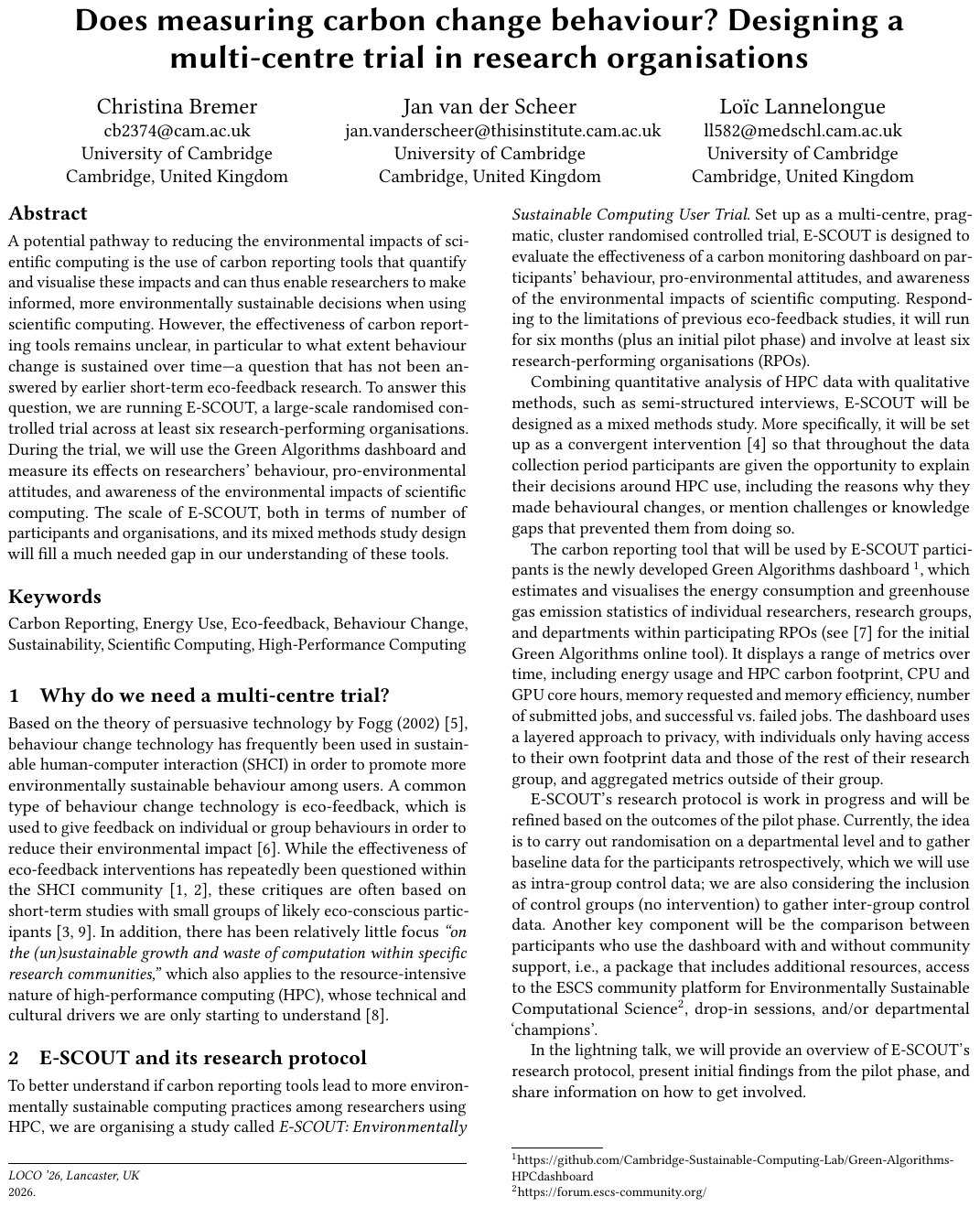}
\includepdf[pages=2-,scale=0.89,pagecommand={\thispagestyle{fancy}}]{L01.pdf}

\fancyfoot[L]{\footnotesize LOCO2026/L02 \quad LOCO 2026 Lightning Talk Abstracts}
\fancyfoot[R]{\footnotesize \thepage}
\includepdf[pages=1,scale=0.89,pagecommand={\thispagestyle{fancy}},addtotoc={1,section,1,{L02: Carbon and energy monitoring dashboard},l02}]{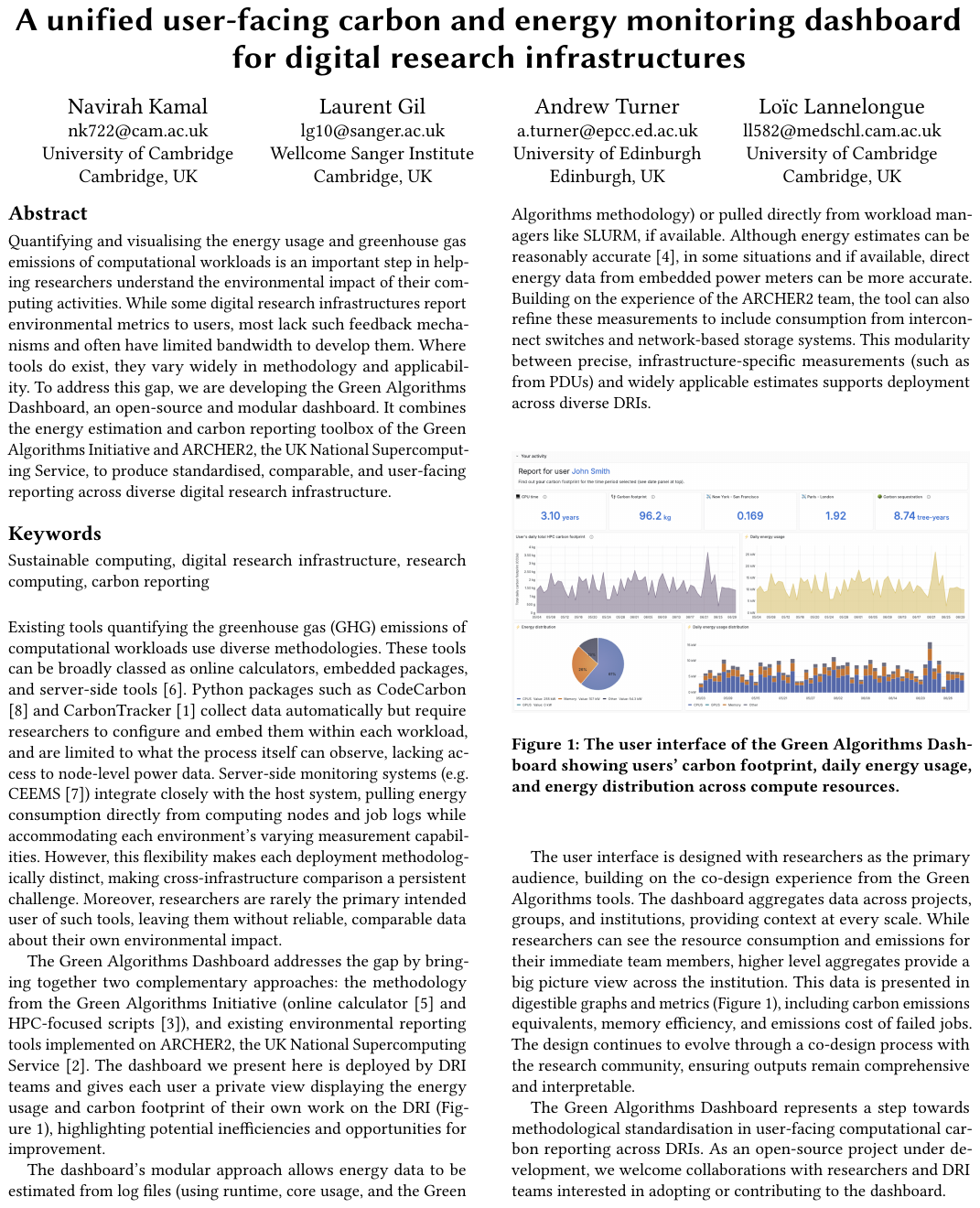}
\includepdf[pages=2-,scale=0.89,pagecommand={\thispagestyle{fancy}}]{L02.pdf}

\fancyfoot[L]{\footnotesize LOCO2026/L03 \quad LOCO 2026 Lightning Talk Abstracts}
\fancyfoot[R]{\footnotesize \thepage}
\includepdf[pages=1,scale=0.89,pagecommand={\thispagestyle{fancy}},addtotoc={1,section,1,{L03: Carbon-Aware Workflow Scheduling Under Deadlines},l03}]{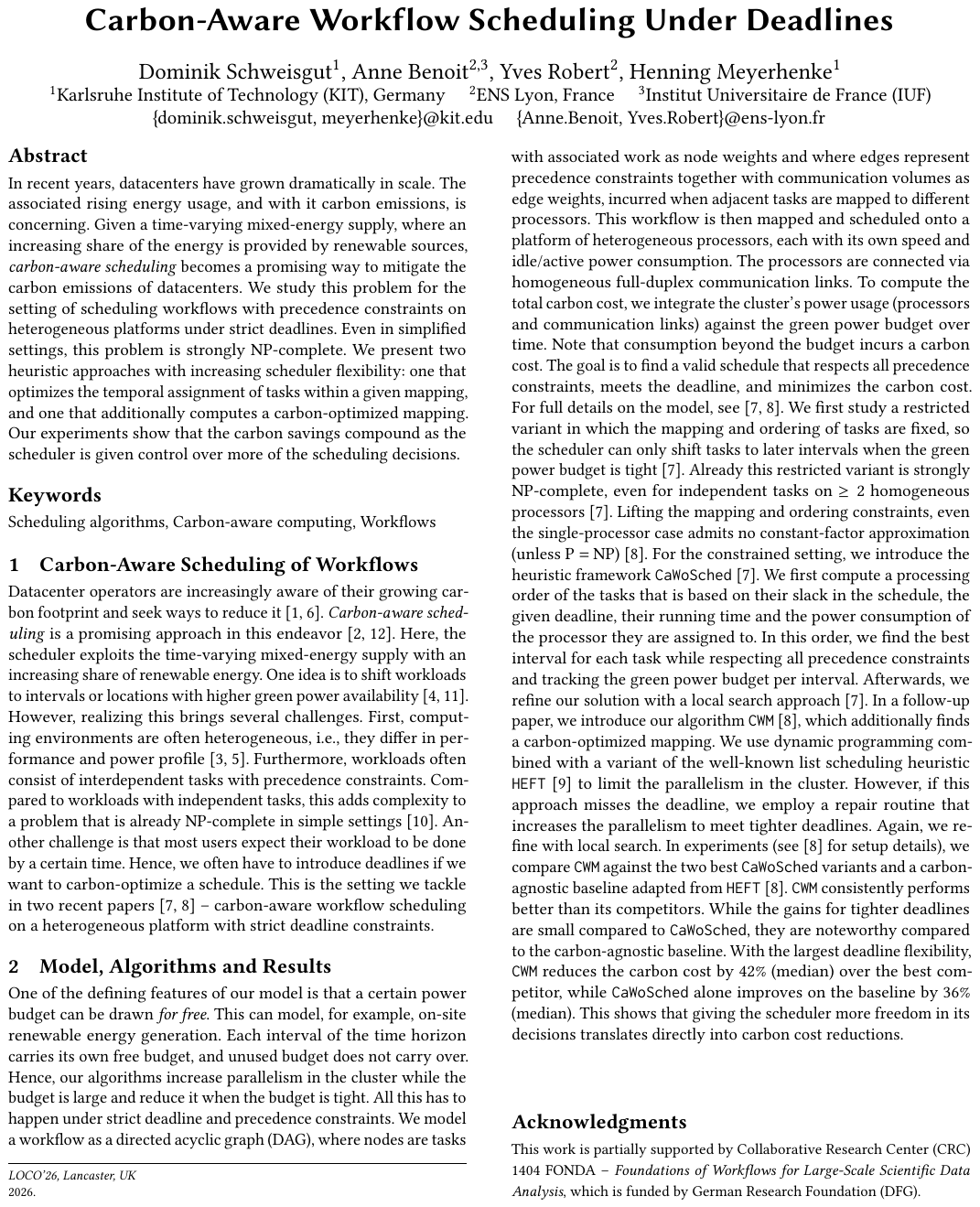}
\includepdf[pages=2-,scale=0.89,pagecommand={\thispagestyle{fancy}}]{L03.pdf}

\fancyfoot[L]{\footnotesize LOCO2026/L04 \quad LOCO 2026 Lightning Talk Abstracts}
\fancyfoot[R]{\footnotesize \thepage}
\includepdf[pages=1,scale=0.89,pagecommand={\thispagestyle{fancy}},addtotoc={1,section,1,{L04: carbon.txt and DIST},l04}]{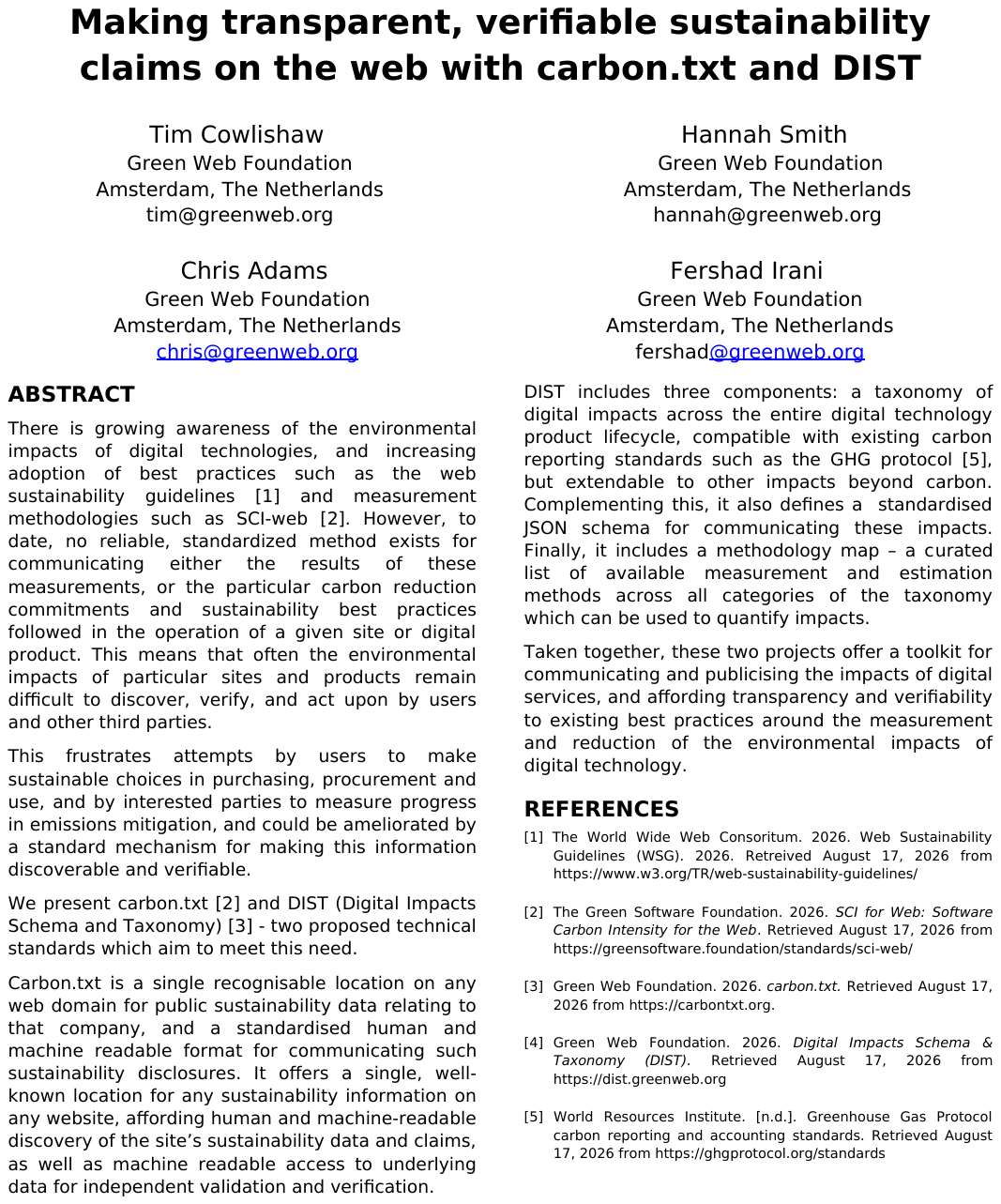}

\fancyfoot[L]{\footnotesize LOCO2026/L05 \quad LOCO 2026 Lightning Talk Abstracts}
\fancyfoot[R]{\footnotesize \thepage}
\includepdf[pages=1,scale=0.89,pagecommand={\thispagestyle{fancy}},addtotoc={1,section,1,{L05: Green BOA},l05}]{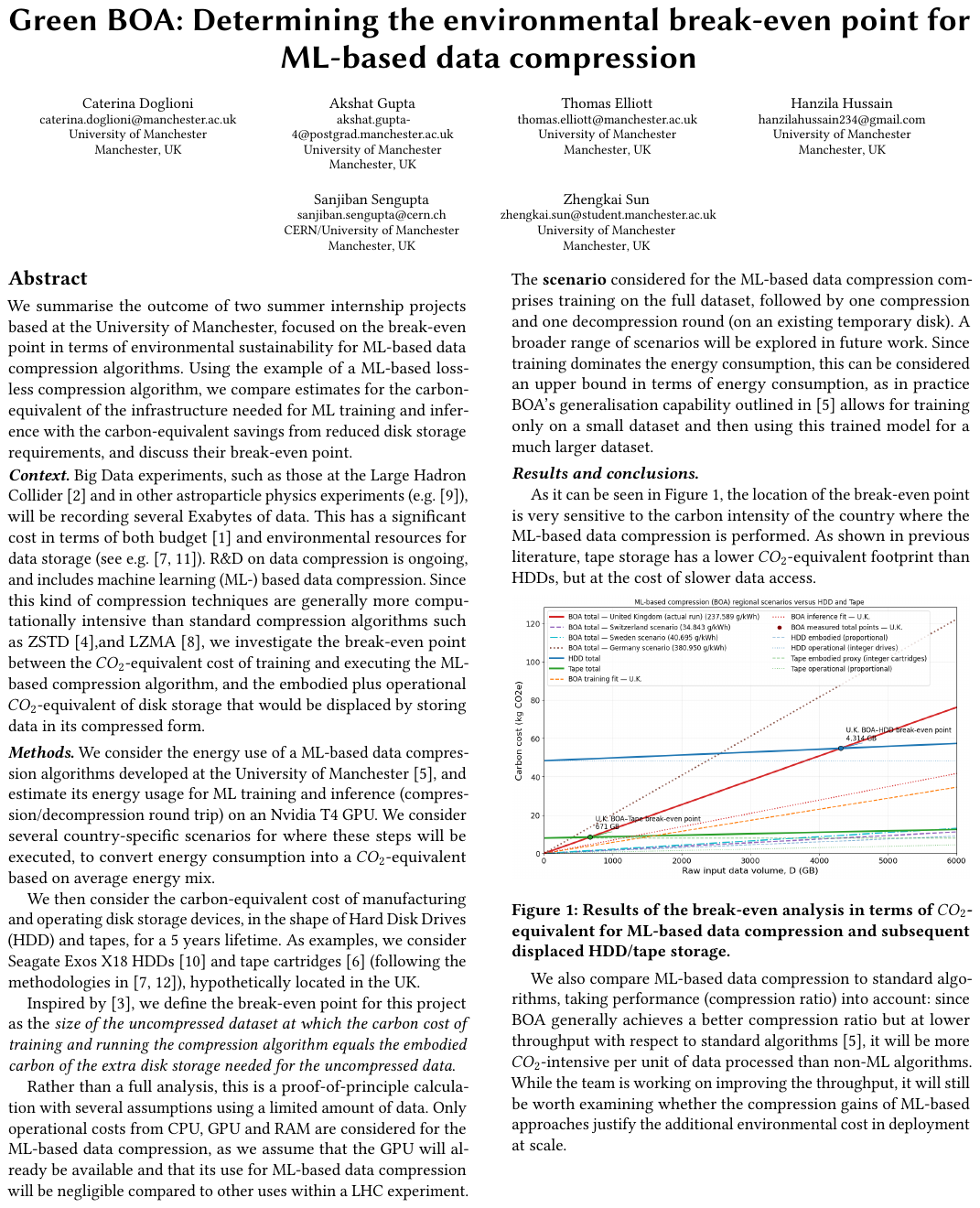}
\includepdf[pages=2-,scale=0.89,pagecommand={\thispagestyle{fancy}}]{L05.pdf}

\fancyfoot[L]{\footnotesize LOCO2026/L06 \quad LOCO 2026 Lightning Talk Abstracts}
\fancyfoot[R]{\footnotesize \thepage}
\includepdf[pages=1,scale=0.89,pagecommand={\thispagestyle{fancy}},addtotoc={1,section,1,{L06: Green TTA},l06}]{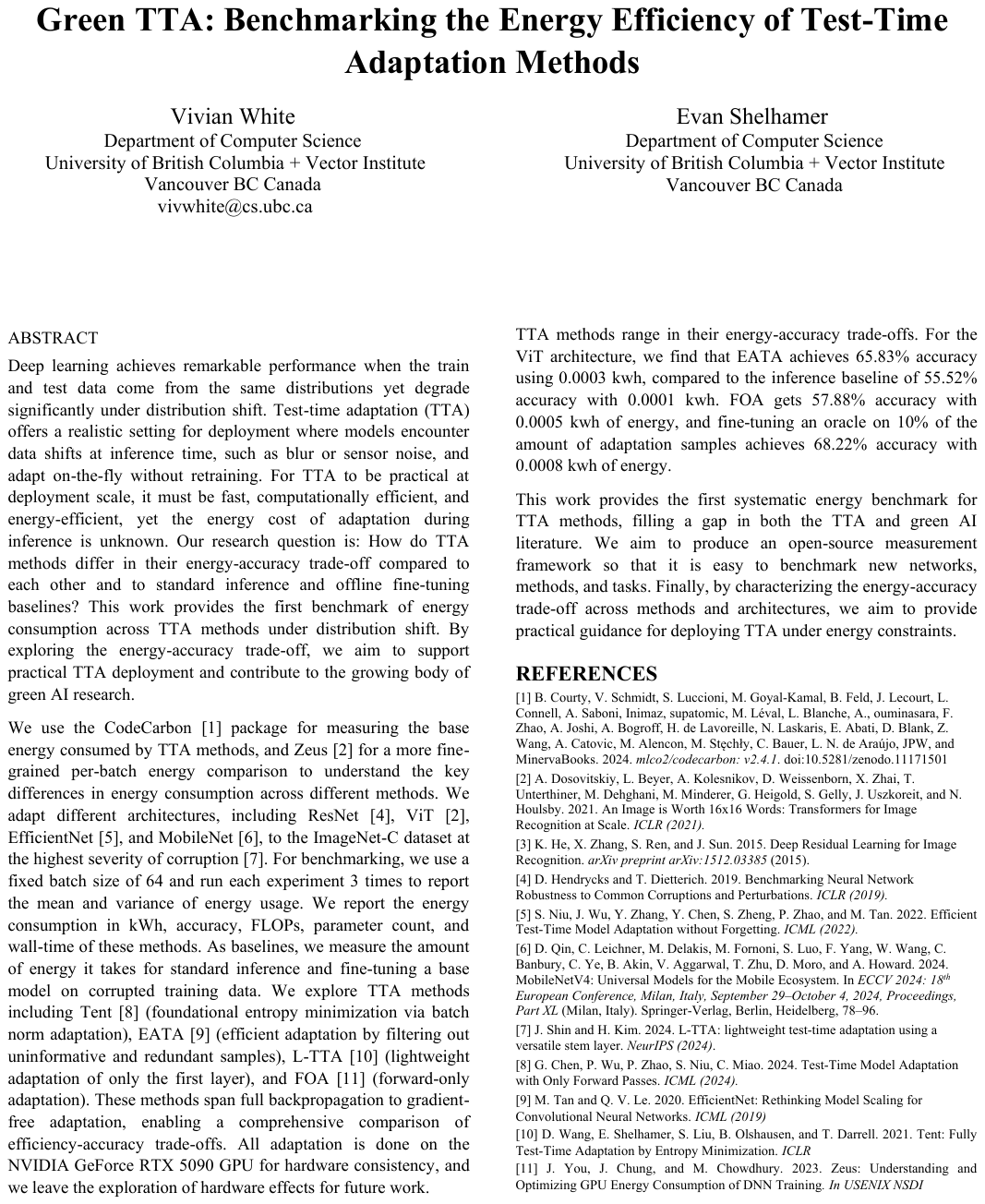}

\fancyfoot[L]{\footnotesize LOCO2026/L07 \quad LOCO 2026 Lightning Talk Abstracts}
\fancyfoot[R]{\footnotesize \thepage}
\includepdf[pages=1,scale=0.89,pagecommand={\thispagestyle{fancy}},addtotoc={1,section,1,{L07: Computational sufficiency in land surface modelling},l07}]{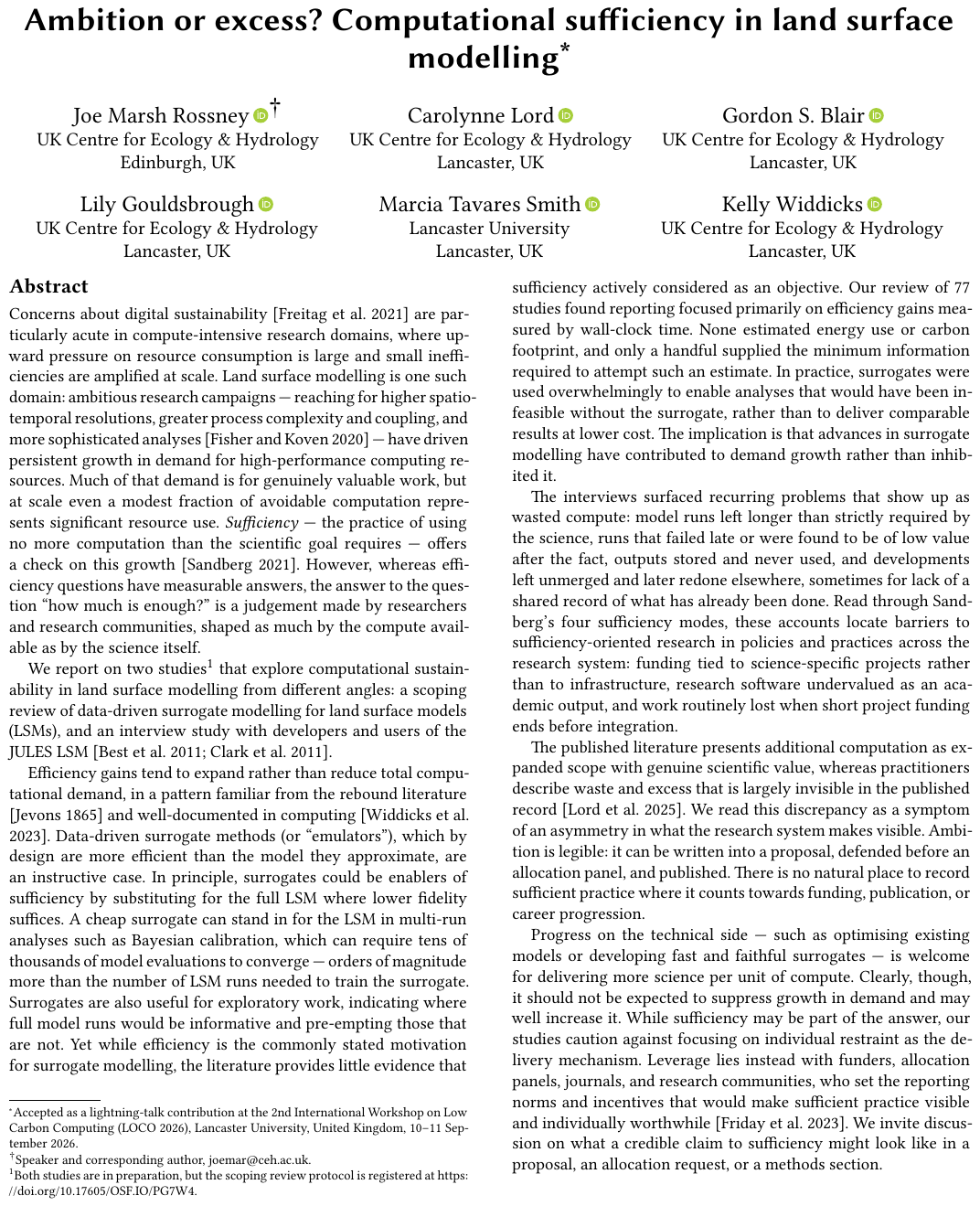}
\includepdf[pages=2-,scale=0.89,pagecommand={\thispagestyle{fancy}}]{L07.pdf}

\fancyfoot[L]{\footnotesize LOCO2026/L08 \quad LOCO 2026 Lightning Talk Abstracts}
\fancyfoot[R]{\footnotesize \thepage}
\includepdf[pages=1,scale=0.89,pagecommand={\thispagestyle{fancy}},addtotoc={1,section,1,{L08: Interoperability as a Carbon Reduction Strategy},l08}]{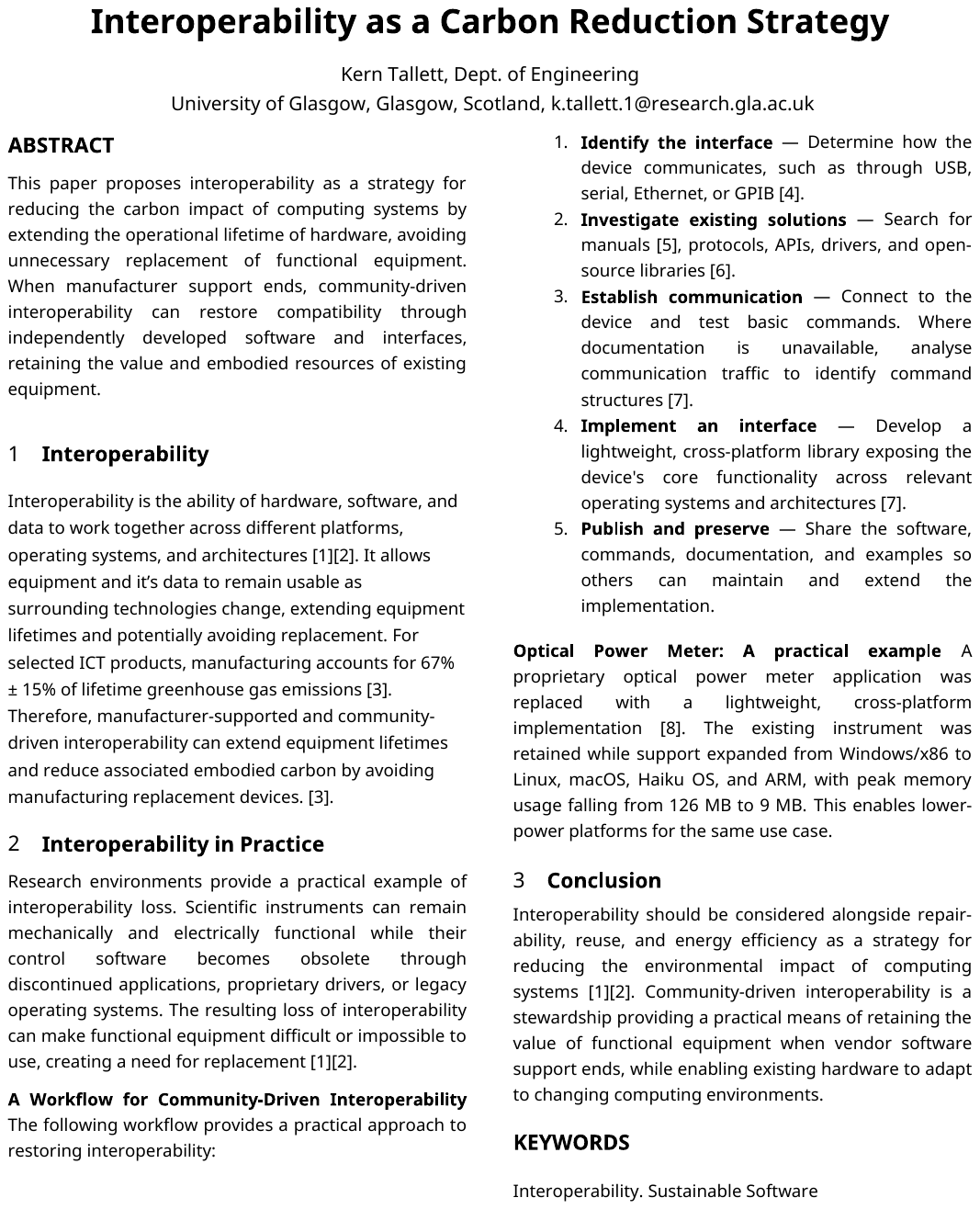}
\includepdf[pages=2-,scale=0.89,pagecommand={\thispagestyle{fancy}}]{L08.pdf}

\fancyfoot[L]{\footnotesize LOCO2026/L09 \quad LOCO 2026 Lightning Talk Abstracts}
\fancyfoot[R]{\footnotesize \thepage}
\includepdf[pages=1,scale=0.89,pagecommand={\thispagestyle{fancy}},addtotoc={1,section,1,{L09: Sustainability-Aware GEMM Benchmarking},l09}]{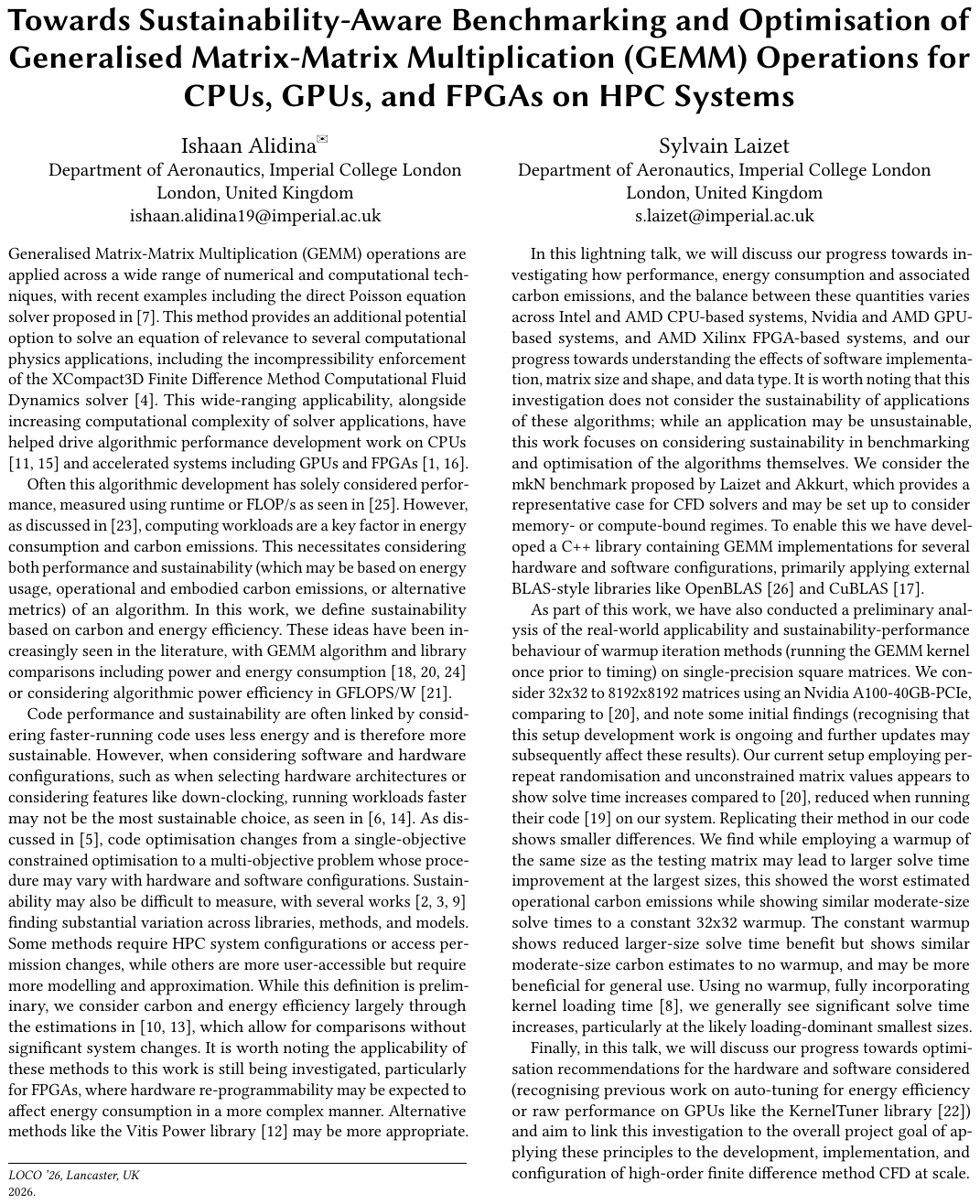}
\includepdf[pages=2-,scale=0.89,pagecommand={\thispagestyle{fancy}}]{L09.pdf}

\fancyfoot[L]{\footnotesize LOCO2026/L10 \quad LOCO 2026 Lightning Talk Abstracts}
\fancyfoot[R]{\footnotesize \thepage}
\includepdf[pages=1,scale=0.89,pagecommand={\thispagestyle{fancy}},addtotoc={1,section,1,{L10: Small-scale solar-powered social media},l10}]{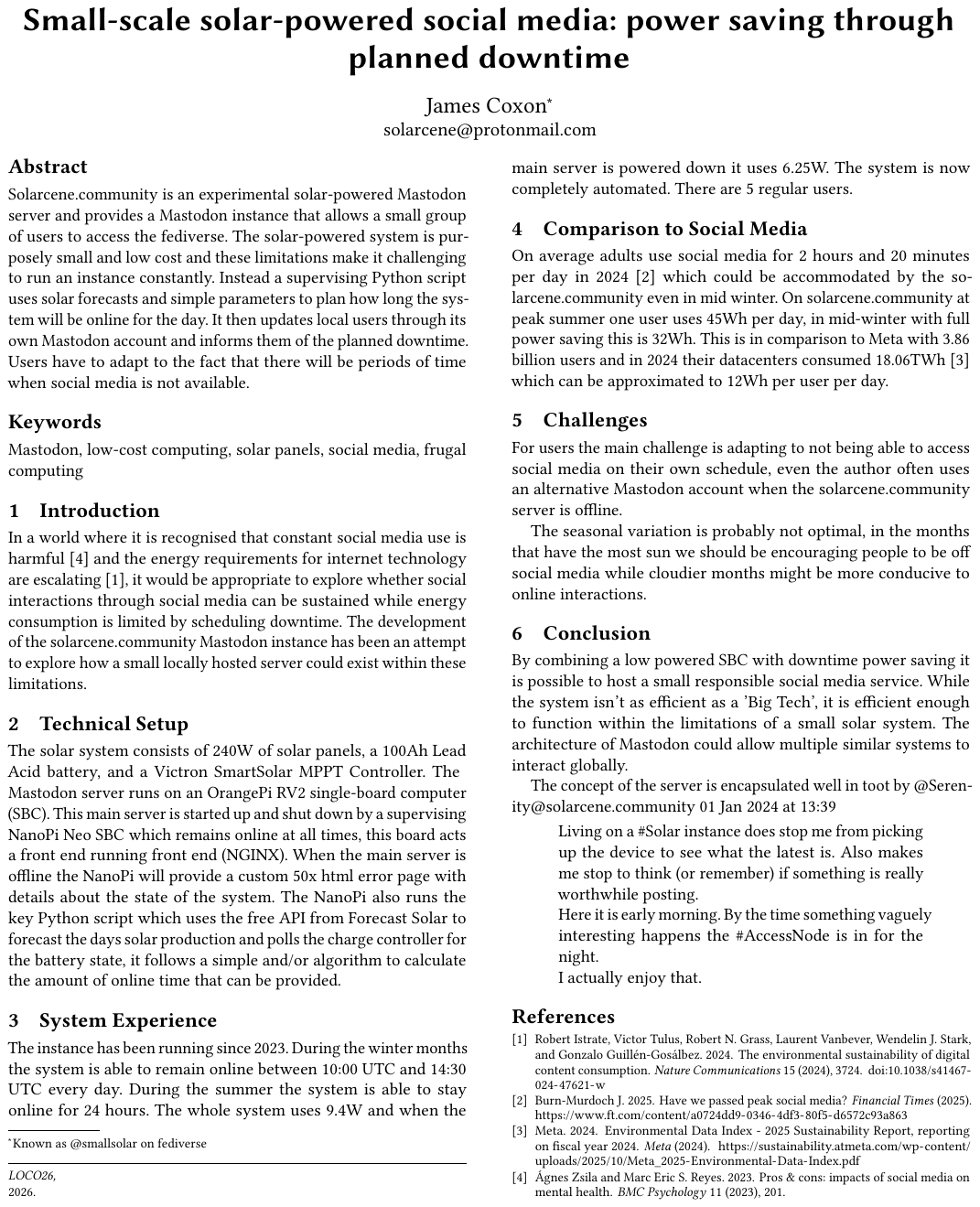}

\fancyfoot[L]{\footnotesize LOCO2026/L11 \quad LOCO 2026 Lightning Talk Abstracts}
\fancyfoot[R]{\footnotesize \thepage}
\includepdf[pages=1,scale=0.89,pagecommand={\thispagestyle{fancy}},addtotoc={1,section,1,{L11: Counting the Stones in my Computer},l11}]{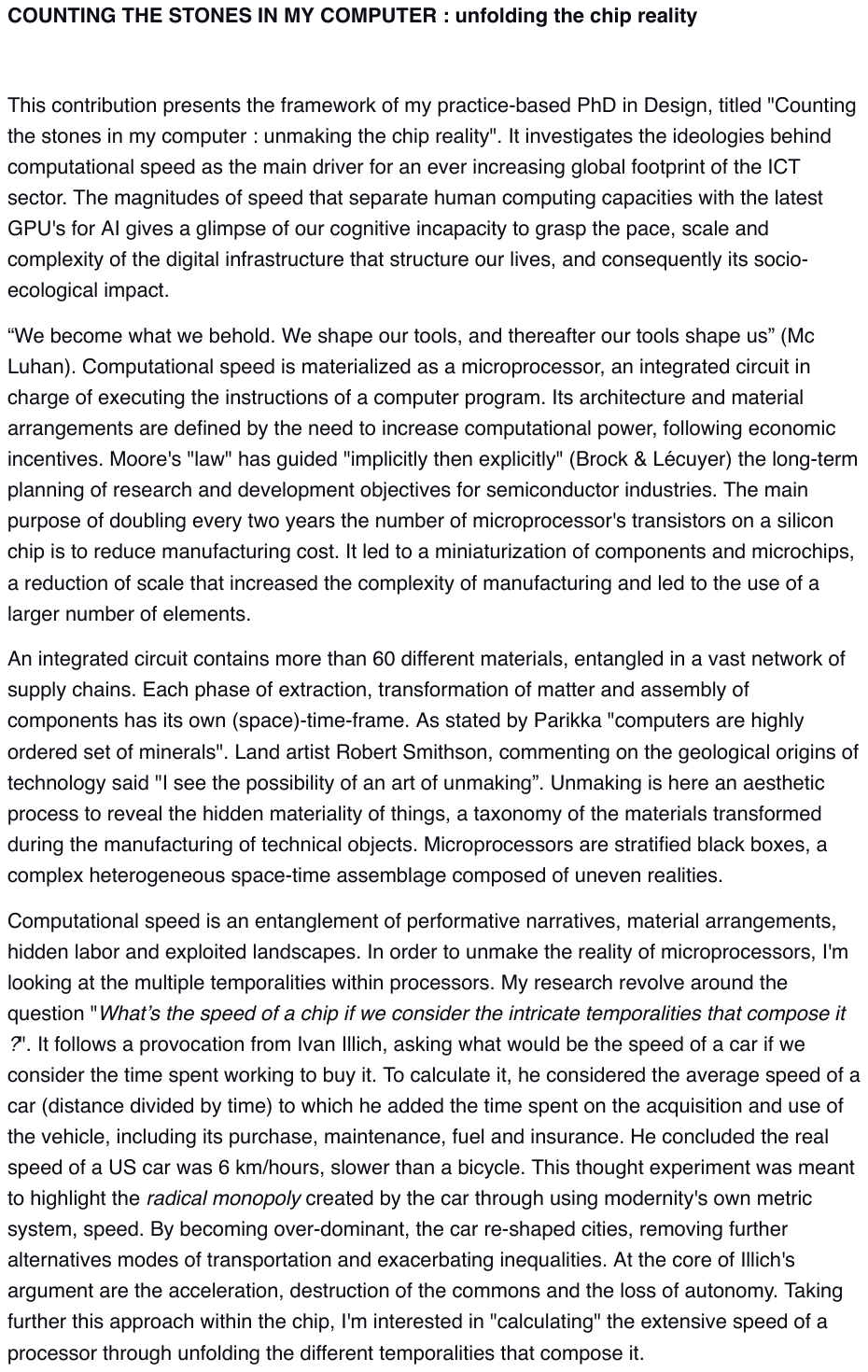}
\includepdf[pages=2-,scale=0.89,pagecommand={\thispagestyle{fancy}}]{L11.pdf}

\fancyfoot[L]{\footnotesize LOCO2026/L12 \quad LOCO 2026 Lightning Talk Abstracts}
\fancyfoot[R]{\footnotesize \thepage}
\includepdf[pages=1,scale=0.89,pagecommand={\thispagestyle{fancy}},addtotoc={1,section,1,{L12: Towards Post-Growth AI},l12}]{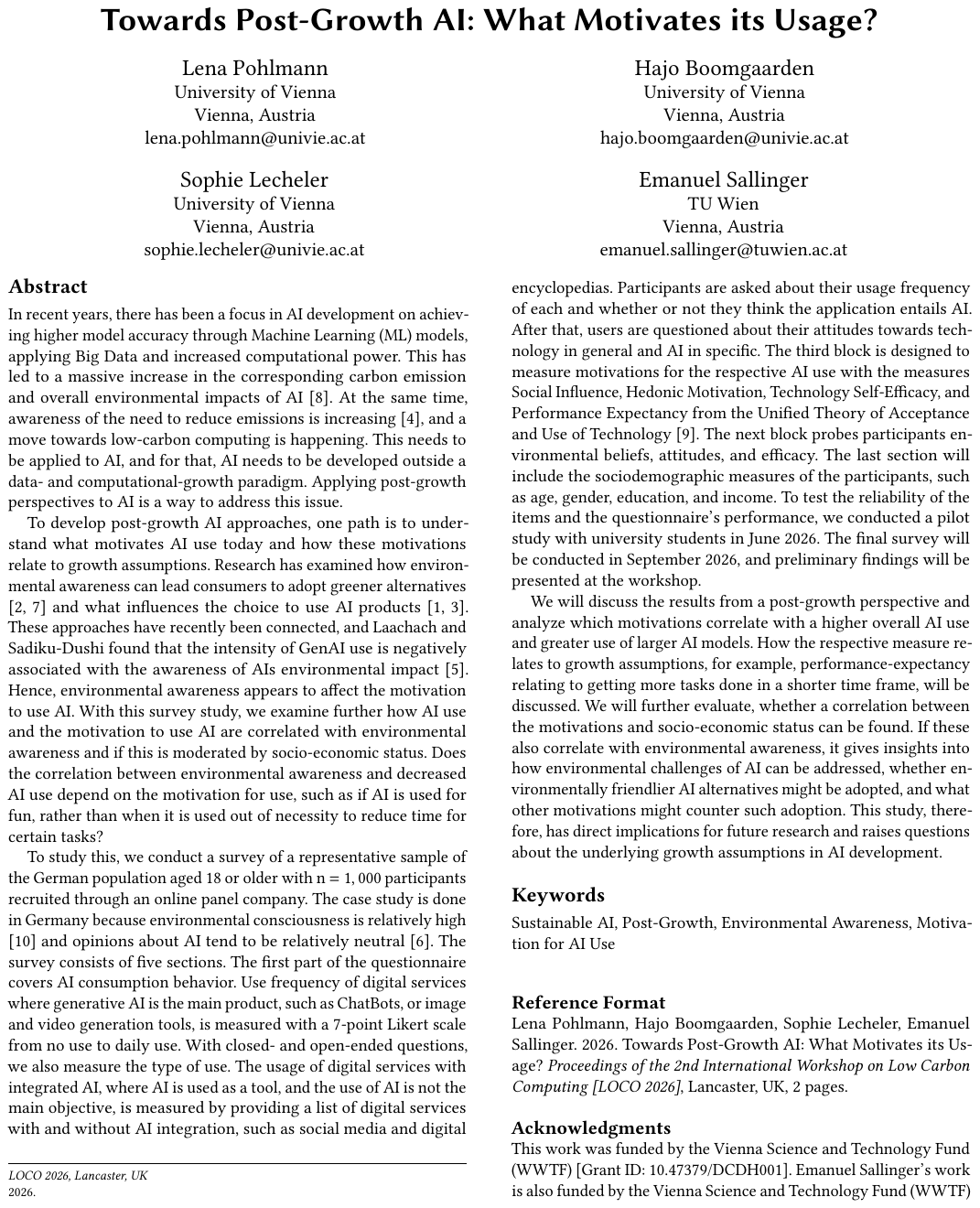}
\includepdf[pages=2-,scale=0.89,pagecommand={\thispagestyle{fancy}}]{L12.pdf}

\fancyfoot[L]{\footnotesize LOCO2026/L13 \quad LOCO 2026 Lightning Talk Abstracts}
\fancyfoot[R]{\footnotesize \thepage}
\includepdf[pages=1,scale=0.89,pagecommand={\thispagestyle{fancy}},addtotoc={1,section,1,{L13: Software-centric measures against premature hardware obsolescence},l13}]{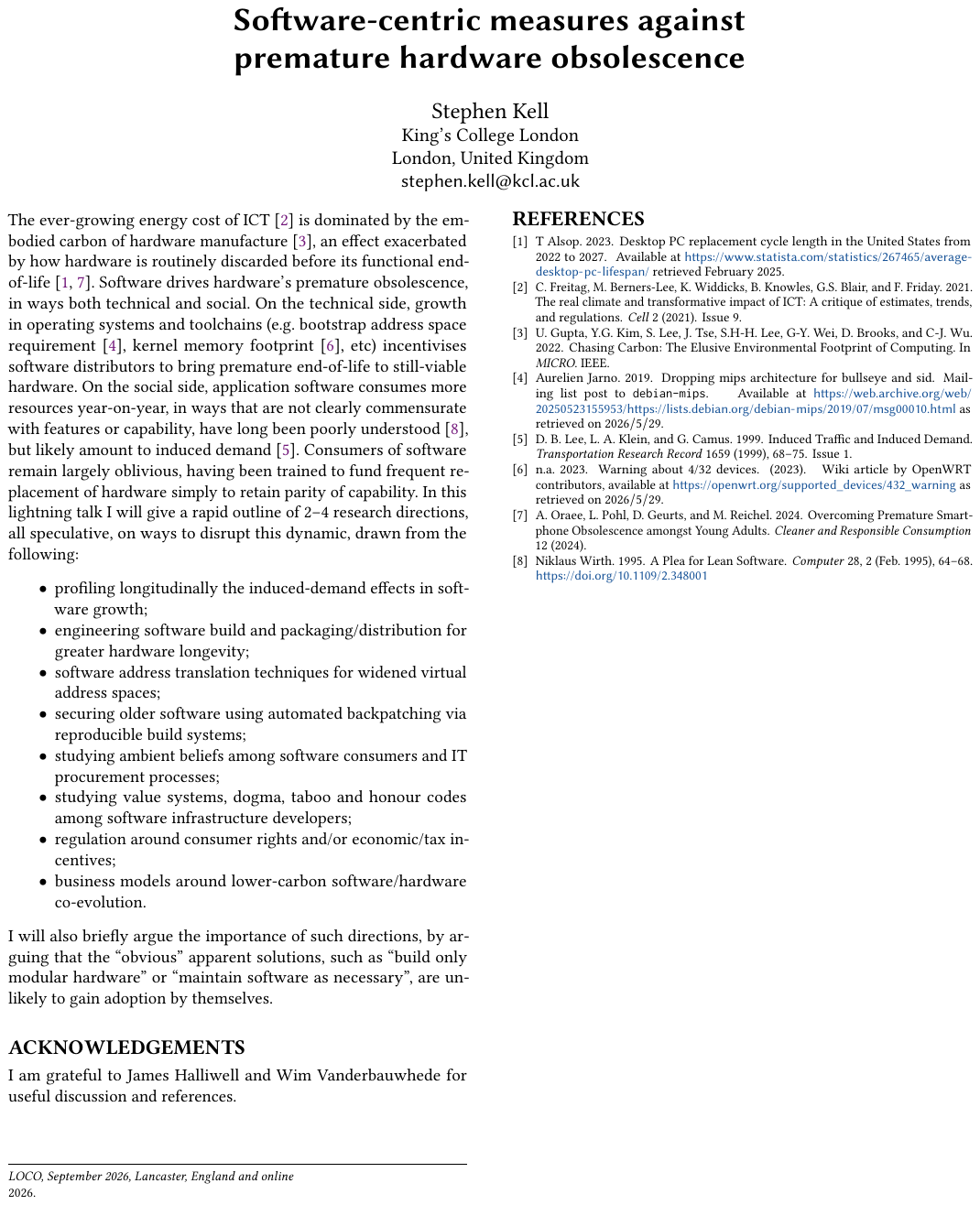}

\fancyfoot[L]{\footnotesize LOCO2026/L14 \quad LOCO 2026 Lightning Talk Abstracts}
\fancyfoot[R]{\footnotesize \thepage}
\includepdf[pages=1,scale=0.89,pagecommand={\thispagestyle{fancy}},addtotoc={1,section,1,{L14: When to Train: A Carbon-Aware Scheduler for ML Workloads},l14}]{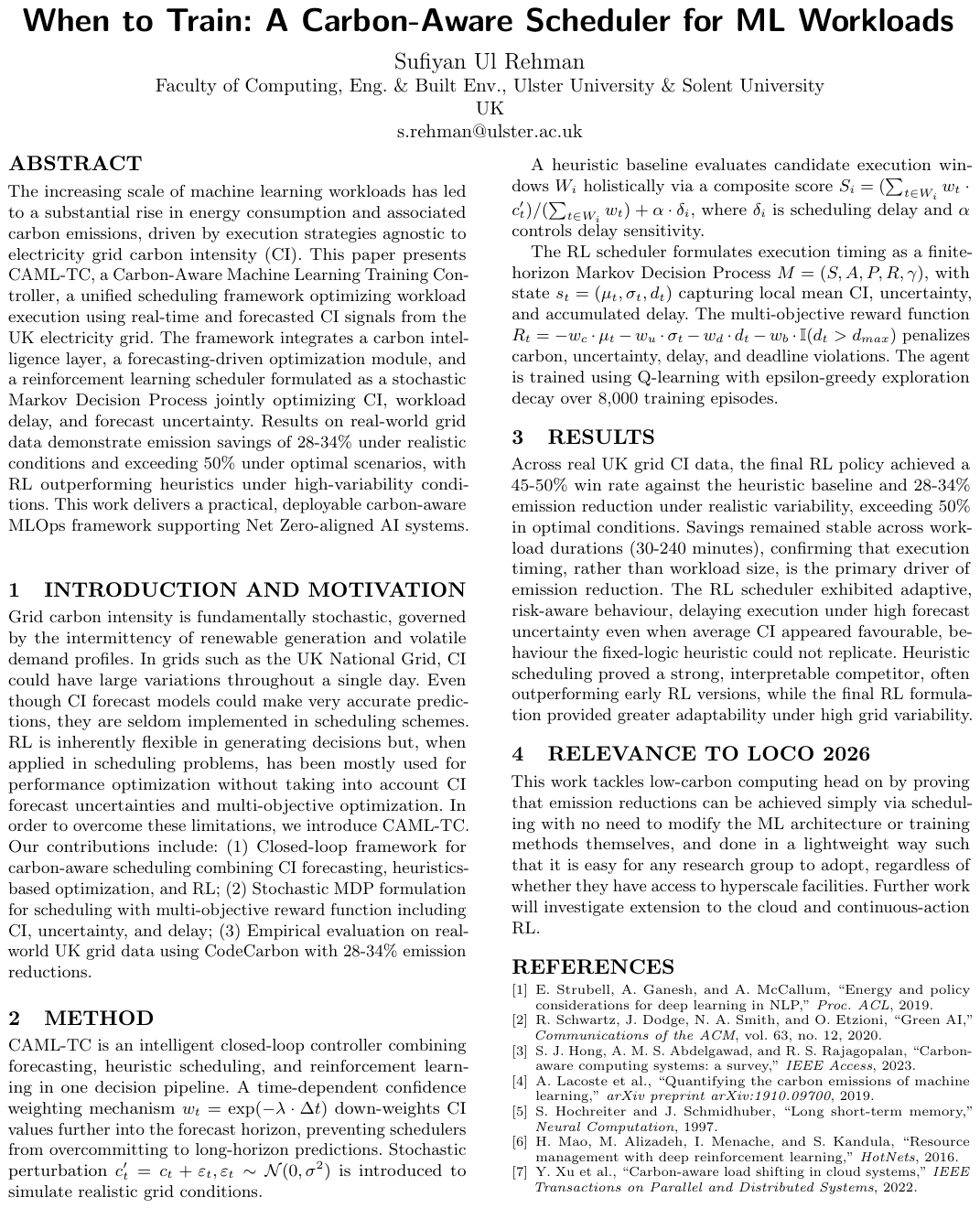}

\end{document}